\documentclass[12pt]{article}

\usepackage{graphicx}
\usepackage{cite}
\usepackage{amsmath}

\begin{document}
\title{\textbf{Observation of the DNA ion-phosphate vibrations}}
\author{L.A. Bulavin$^{1}$, S.N. Volkov$^{2}$,
S.Yu. Kutovy$^{1}$, S.M.
Perepelytsya$^{1}$\\
\small $^{1}$Taras Shevchenko National University, Department of
Physics,\\\small
64 Volodymyrska St., Kiev, 01033, Ukraine\\
\small $^{2}$Bogolyubov Institute for Theoretical Physics, NAS of
Ukraine,\\\small 14-b Metrologichna St., Kiev, 03680, Ukraine}
\maketitle

 \begin{abstract}

The low-frequency Raman spectra of Na-and Cs-DNA water solutions
have been studied to determine the mode of counterion vibrations
with respect to phosphate groups of the DNA double helix. The
obtained spectra are characterized by the water band near 180
cm$^{-1}$ and by several DNA bands near 100 cm$^{-1}$. The main
difference between Na- and Cs-DNA spectra is observed in case of
the band 100 cm$^{-1}$. In Cs-DNA spectra this band has about
twice higher intensity than in Na-DNA spectra. The comparison of
obtained spectra with the calculated frequencies of Na- and Cs-DNA
conformational vibrations [\emph{Perepelytsya S.M., Volkov S.N.}
Eur. Phys. J. E. \textbf{24}, 261 (2007)] show that the band 100
cm$^{-1}$ in the spectra of Cs-DNA is formed by the modes of both
H-bond stretching vibrations and vibrations of caesium
counterions, while in Na-DNA spectra the band 100 cm$^{-1}$ is
formed by the mode of H-bond stretching vibrations only. The modes
of sodium counterion vibrations have a frequency 180 cm$^{-1}$,
and they do not rise above the water band. Thus, the increase in
intensity of the band 100 cm$^{-1}$ in the spectra of Cs-DNA as
compared with Na-DNA is caused by the mode of ion-phosphate vibrations.
\end{abstract}

Under the natural conditions the structure of DNA macromolecule is
stabilized by alkali metal counterions that neutralize the
negatively charged phosphate groups of the double helix backbone
~\cite{1}. The counterions, bonded to the phosphate groups, form a
regular structure along the DNA backbone. This structure may be
considered as an ion lattice. The formation and existence of such
lattice should be characterized by the specific ion-phosphate
vibrations. Therefore, the purpose of this work is to make an
experimental observation of the DNA ion-phosphate mode, proving
the existence of the ion-phosphate lattice.

The earlier calculations for DNA with alkali metal counterions
~\cite{2,3,4} have been shown that the mode of ion-phosphate
vibrations must be in the low-frequency spectra range ($<$250
cm$^{-1}$). The calculated frequency of ion-phosphate vibrations
in case of Na$^{+}$, K$^{+}$, Rb$^{+}$ and Cs$^{+}$ counterions
decreases from 180 to 90 cm$^{-1}$ as counterion mass increases
~\cite{3,4}. The calculated amplitudes of vibrations have been
shown that the character of DNA conformational vibrations is
different in case of light (Na$^{+}$ and K$^{+}$) and heavy
(Rb$^{+}$ and Cs$^{+}$) counterions ~\cite{4}. This difference is
expected to be seen in the experimental spectra.

Existing experimental data has been shown that in this range of
nucleic acid spectra there are modes, depending on the counterion
type and concentration ~\cite{5,6}. In the spectra of
poly(rI)$\cdot$poly(rC) dry films a mode depending on the type of
alkali metal counterion has been observed ~\cite{5}. The frequency
of this mode decreases from 150 to 110 and 95 cm$^{-1}$ after
substitution of K$^{+}$ counterions for Rb$^{+}$ and Cs$^{+}$,
respectively. In the spectra of poly(dA)$\cdot$poly(dT) dry films
the mode 170 cm$^{-1}$ has been observed ~\cite{6}. The intensity
of this mode reduces as concentration of Na$^{+}$ counterion
decreases. However, in the DNA low-frequency spectra there are
modes, characterizing internal vibrations of the double helix,
such as vibrations of H-bond stretching in the base pairs and
deformation of sugar rings. These modes are situated in the
frequency range from 60 to 120 cm$^{-1}$ ~\cite{6,7,8,9,10}.

For the experimental determination of the ion-phosphate mode among
the modes of the DNA low-frequency spectra, the dependence of this
mode on counterion type should be used. Therefore, in this work
the low-frequency Raman spectra of DNA with light (Na$^{+}$) and
heavy (Cs$^{+}$) counterions have been studied. As a result, the
mode of ion-phosphate vibrations is observed in the spectra of
Cs-DNA water solutions.

The samples of Na- and Cs-DNA are prepared using the DNA sodium
salt from calf thymus with molecular weight of 10$^{7}$ a.u.m. The
dry DNA is dissolved in distilled water to concentration of 0,2 \%
(by weight) and then is treated by ultrasound with the frequency
of 22 kHz. The molecular weight of DNA after ultrasound treatment
is determined using electrophoresis in agarous gel. The results
show that molecular weight of DNA decreases to 10$^{5}$ a.u.m.
Cs-DNA is obtained from Na-DNA, by replacing Na$^{+}$ for Cs$^{+}$
with the use of methods described in ~\cite{7}. To perform the
counterion exchange the salt CsCl (0.5 M) and two volumes of
ethanol are added to 0.2 \% solution of Na-DNA. In the alcohol
solution, containing large concentration of CsCl salt, Na$^{+}$
counterions of DNA are replaced by Cs$^{+}$ counterions. The
resulting mixture has been stored under temperature -20 $^{0}$C
during tree days. The DNA macromolecule precipitates. The
counterion exchange is controlled with the help of Auger spectra
of obtained samples. To remove an excess salt from the sample the
procedure is repeated once again with much smaller concentration
of CsCl salt (0.05 M). The obtained sediment has been dried under
the ambient temperature during three days. The dry Cs-DNA is
dissolved in water to the concentration 4\%. The concentration of
DNA in obtained water solutions is controlled by absorbtion band
at 256 nm. Some additional salt (NaCl or CsCl) with the
concentraion 0.5 M is added to the obtained solutions of Na- and
Cs-DNA (4\%).

The samples are excited by argon laser beam with wave-length 5145
\AA. The real power of radiation transmitted to the sample, is
about 60$\div$80 mW. The track of laser beam, passing through the
cuvette with the sample, is parallel to the slot of monochromator
(DFS-24). The volume and hight of cuvette is about 0.06 ml and 15
mm, respectively. The studied DNA low-frequency Raman spectra
being less intensive, in order to increase a signal/noise ratio,
the exposition time in the point is increased to 4$\div$8 s. As a
result, the spectra of Na- and Cs-DNA water solutions are obtained
within frequency range 30$\div$230 cm$^{-1}$ (Fig. 1a). The DNA
low-frequency spectra are recorded to have characteristic band
near 100 cm$^{-1}$. The intensity of this band in the spectra of
Cs-DNA is higher than in the spectra of Na-DNA.

\begin{figure}[t!]
\includegraphics[width=17cm]{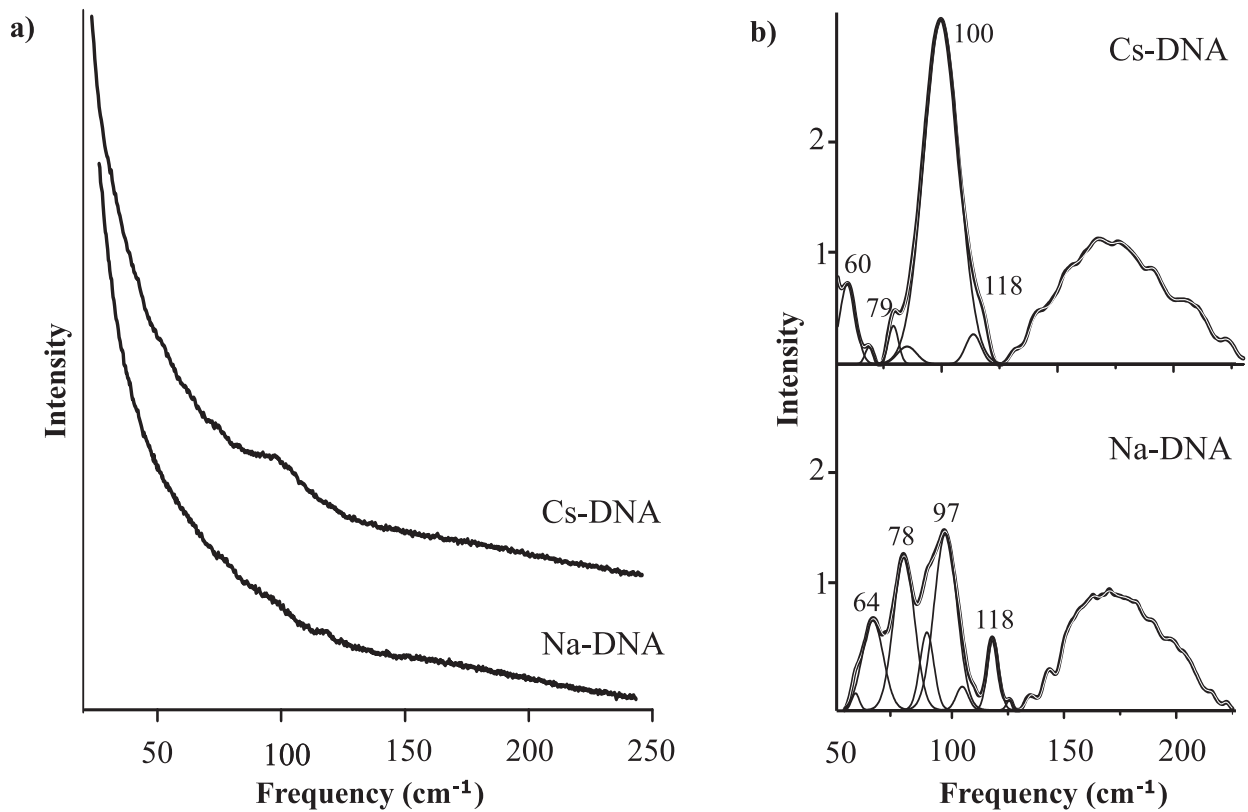}
\caption{The low-frequency Raman spectra of Na- and Cs-DNA water
solutions. The spectra before (a) and after (b) mathematical
treatment.} \label{Fi:1}
\end{figure}

To find the mode of DNA ion-phosphate vibrations, the detailed
mathematical treatment of the spectra has been made. The spectra
shapes are fitted by Lorenz curve. After substracting the spectra
from Lorenz curve, the resulted spectra are expanded into Gaussian
curves. As a result, the mode frequencies are determined with
accuracy $\pm$2 cm$^{-1}$. The intensities of the vibrational
modes are normalized by the intensity of broad band near 180
cm$^{-1}$, since the intensity of this band has a similar value in
both spectra. As it is known, the band 180 cm$^{-1}$ characterizes
translation vibrations of water molecules ~\cite{11}. The
normalized spectra of Na- and Cs-DNA are shown in figure 1b.

As one can see, within the frequency range 50$\div$120 cm$^{-1}$
of Na-DNA spectra the modes 78, 118 cm$^{-1}$ and the bands 60, 97
cm$^{-1}$ are registered. In the spectra of Cs-DNA, within the
same frequency range, the modes 60, 79, 118 cm$^{-1}$ and
intensive band 100 cm$^{-1}$ are observed. The intensity of band
100 cm$^{-1}$ in the spectra of Cs-DNA is about twice as much than
in the spectra of Na-DNA. Unlikely, modes 79 and 118 cm$^{-1}$ are
much intensive in the spectra of Na-DNA than in the spectra of
Cs-DNA. At higher frequencies (more than 130 cm$^{-1}$) the broad
band 180 cm$^{-1}$ is observed both in the Na- and Cs-DNA spectra.
The analysis of this band for selecting the DNA vibrational modes
is a subject of special research, therefore, it is not discussed
herein.

To determine the mode of DNA ion-phosphate vibrations the obtained
frequencies of vibrations are compared with the existing
experimental and theoretical data ~\cite{2,3,4,5,6,7,8,9}. The
frequency values and the character of DNA structural motions for
observed modes are shown in table 1. The structure of obtained
spectra for Na-DNA is seen to be similar to the spectra for
polynucleotide dry films and for DNA crystals ~\cite{6,8}. The
spectra of Cs-DNA has been experimentally investigated
insufficiently, however, it is known that at 95 cm$^{-1}$ the mode
depending on the counterion type is observed ~\cite{5}.

The comparison of experimental data (Fig. 1b) with the
calculations ~\cite{2,3,4}, has shown that the band 64 cm$^{-1}$
in the spectra of Na-DNA characterizes the vibrations of the bases
with respect to the phosphate groups of the DNA backbone
($\omega^{+}_{H+S}$). These vibrations cause the deformation of
sugar rings and stretching of H-bonds in the base pairs. The band
97 cm$^{-1}$ characterizes the vibrations of bases, which cause
the stretching of H-bonds ($\omega^{+}_{H}$). According to
calculations ~\cite{4}, the peak 79 cm$^{-1}$ in the spectra of
Na-DNA characterizes the vibrations of bases with respect to the
phosphate groups of the DNA backbone ($\omega^{+}_{S}$). These
vibrations cause the deformations of sugar rings and occur without
stretching of H-bonds in the base pairs. The intensity of mode 118
cm$^{-1}$ changes after substitution of Na$^{+}$ for Cs$^{+}$
similarly to the mode 78 cm$^{-1}$, therefore it may characterize
the deformation of sugar ring. According to calculations
~\cite{4}, the modes of Na-DNA ion-phosphate vibrations are
located inside the band 180 cm$^{-1}$ that characterizes the
vibrations of water molecules.

\begin{center}
 \textbf{Table 1.} The frequencies of  Na- and
Cs-DNA vibrational spectra (cm$^{-1}$). \emph{sh} -- shoulder;
\emph{b} -- band. $\omega_{ion}^{+}$ and $\omega_{ion}^{-}$   are
the frequencies of ion-phosphate vibrations; $\omega_{H}^{+}$ and
$\omega_{H+S}^{+}$ are the frequencies of H-bond stretching in
base pairs; $\omega_{S}^{-}$ is the frequency of sugar ring
deformations.
\begin{tabular}{lccccc}\hline
\textbf{Na-DNA} &&&&&\\
Our experiment(Fig. 1a)&57\emph{sh};64&78&88\emph{sh};97\emph{b};110\emph{sh}&118&180\emph{b}\\
Experiment ~\cite{6}&63&80&95;106&--&170\\
Experiment ~\cite{8}&68&--&96&120&--\\
Theory
~\cite{4}&58$\omega_{H+S}^{+}$&79$\omega_{S}^{-}$&110$\omega_{H}^{+}$&--&181$\omega_{ion}^{\pm}$\\\hline
\textbf{Cs-DNA} &&&&&\\
Our experiment
(Fig. 1b)&--&60;69\emph{sh};79&85\emph{sh};100\emph{b};113\emph{sh}&118\emph{sh}&180\emph{b}\\
Experiment ~\cite{5}&--&--&95&--&--\\
Theory
~\cite{4}&44$\omega_{H+S}^{+}$;&60$\omega_{S}^{-}$&94$\omega_{H}^{+}$;103$\omega_{ion}^{-}$;110$\omega_{ion}^{+}$&--&--\\\hline
  \end{tabular}
\end{center}

The calculations for Cs-DNA ~\cite{3,4} show that the peak 60
cm$^{-1}$ characterizes the vibrations of bases with respect to
the phosphate backbone of DNA, causing the deformations of sugar
rings ($\omega^{+}_{S}$). The band 100 cm$^{-1}$ is formed by the
modes of H-bond stretching vibrations ($\omega_{H}^{+}$) and
ion-phosphate vibrations. In the spectra of Cs-DNA the mode
$\omega^{+}_{H+S}$, characterizing the vibrations of H-bond
stretching and deformations of sugar rings, should be at 44
cm$^{-1}$ ~\cite{2,3,4}. This frequency is beyond the studied
spectra range. The intensities of modes 79 and 118 cm$^{-1}$ in
the spectra of Cs-DNA are much smaller than in the spectra of
Na-DNA. These modes appear in the spectra of Cs-DNA, since in the
samples of Cs-DNA there are residual Na$^{+}$ ions remaining after
the counterion exchange.

The analysis of obtained experimental data shows that in the
spectra of Cs-DNA the modes of ion-phosphate vibrations and the
modes of H-bond stretching in the base pairs form the band 100
cm$^{-1}$. The intensity of this band in the spectra of Cs-DNA is
about twice higher than in the spectra of Na-DNA (Fig. 1b). The
modes of Na-DNA ion-phosphate vibrations are situated at higher
frequencies (180 cm$^{-1}$), and they are not separated from the
wide water band. To explain why not Na-DNA, but Cs-DNA
ion-phosphate modes are observed the ratio between mode
intensities in the spectra of Na and Cs-DNA should be studied.

\begin{figure}[t!]
\includegraphics[width=17cm]{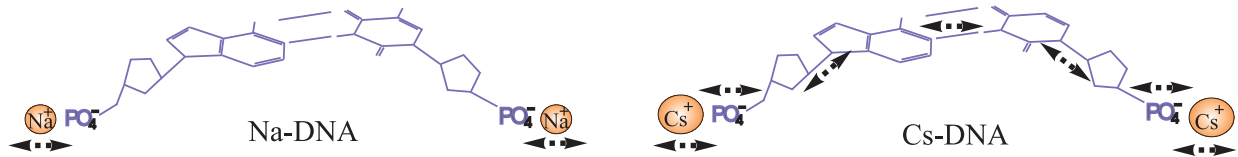}
\caption{The displacements of the structural elements in base pair
in case of ion-phosphate vibrational mode. The arrows indicate the
directions of atomic group displacements. The diagrams were built
using the calculation data ~\cite{4}}\label{Fi:2}
\end{figure}

As it is known the intensity of Raman active mode is proportional
to squared polarizability derivative with respect to normal
coordinate ~\cite{12}. The calculations ~\cite{4} show that Na-DNA
ion-phosphate modes are characterized by vibrations of sodium
counterions solely. In the case of Cs-DNA these modes are
characterized by large displacements of all structure elements of
the double helix (Fig. 2). Hence, in the case of Cs$^{+}$
counterions the displacements of the DNA structural elements
should cause large changes of polarizability of the whole system.
Therefore, the intensity of ion-phosphate modes in the spectra of
Cs-DNA has to be much higher than in the spectra of Na-DNA. This
fact explains an increase in intensity of the band 100 cm$^{-1}$
in the spectra of Cs-DNA as compared with Na-DNA.

To summarize the results of this study it should be noted that the
mode of ion-phosphate vibrations is observed in Cs-DNA Raman
spectra for the first time. This mode takes place within the band
100 cm$^{-1}$, intensifying this band about twice. The observation
of the mode of counterion vibrations with respect to the phosphate
groups of the double helix verifies the existence of the DNA
ion-phosphate lattice that is important for understanding of the
mechanisms of DNA biological functioning.\\

\textbf{Acknowledgements}\\
We would like to thank Prof. F.I. Tovkach for help in preparation
of DNA samples.

\end{document}